\documentclass[11pt]{article} 

\usepackage[utf8]{inputenc} 

\usepackage{geometry} 
\usepackage{graphicx} 

\usepackage{booktabs} 
\usepackage{array} 
\usepackage{paralist} 
\usepackage{verbatim} 
\usepackage{subfig} 
\usepackage{amsmath} 
\usepackage{amssymb} 
\usepackage{epstopdf} 
\usepackage{color}

\usepackage{fancyhdr} 
\usepackage{sectsty}
\allsectionsfont{\sffamily\mdseries\upshape} 

\usepackage[nottoc,notlof,notlot]{tocbibind} 
\usepackage[titles,subfigure]{tocloft} 

\title{Minimum Dominating Sets in Scale-Free Network Ensembles}
\author{F. Moln\'{a}r Jr.$^{1,2}$, S. Sreenivasan$^{2,3}$, B.K. Szymanski$^{2,3}$, G. Korniss$^{1,2,}\footnote{E-mail: korniss@rpi.edu}$}

\begin{document}
\date{April 2, 2013}
\maketitle

\begin{flushleft}
$^{\bf{1}}$ Department of Physics, Applied Physics, and Astronomy, Rensselaer Polytechnic Institute,
110 8$^{th}$ Street, Troy, NY, 12180-3590 USA \\
$^{\bf{2}}$ Social Cognitive Networks Academic Research Center,
Rensselaer Polytechnic Institute, 110 8$^{th}$ Street, Troy, NY, 12180-3590 USA \\
$^{\bf{3}}$ Department of Computer Science,
Rensselaer Polytechnic Institute, 110 8$^{th}$ Street, Troy, NY, 12180-3590 USA \\
\end{flushleft}

\section*{Abstract}

We study the scaling behavior of the size of minimum dominating set
(MDS) in scale-free networks, with respect to network size $N$ and
power-law exponent $\gamma$, while keeping the average degree fixed.
We study ensembles generated by three different network construction
methods, and we use a greedy algorithm to approximate the MDS. With
a structural cutoff imposed on the maximal degree
($k_{\max}=\sqrt{N}$) we find linear scaling of the MDS size with
respect to $N$ in all three network classes. Without any cutoff
($k_{\max}=N-1$) two of the network classes display a transition at
$\gamma \approx 1.9$, with linear scaling above, and vanishingly
weak dependence below, but in the third network class we find linear
scaling irrespective of $\gamma$. We find that the partial MDS,
which dominates a given $z<1$ fraction of nodes, displays
essentially the same scaling behavior as the MDS.

\section*{Introduction}

A central issue arising in the context of networked systems is the
ability to efficiently control, track, or detect the behavior of the
constituent nodes of a network. In static networks or slowly
evolving networks, a solution to this problem often involves
computing a {\it dominating set} of the network. A dominating set of
a network (graph) $\mathcal{G}$ with node set $V$ is a subset of
nodes $S \subseteq V$ such that every node not in $S$ is adjacent to
at least one node in $S$. Example problems in whose solution the
dominating set (or some variant of it) has been shown to play a part
include optimal sensor placement for disease outbreak detection
\cite{Eubank}, controllability of networks \cite{Cowan} and social
influence propagation \cite{Kelleher_1988,Wang}. The smallest dominating set of
$\mathcal{G}$ constitutes its {\it minimum dominating set} (MDS).
Thus, the MDS of a network is the {\em smallest} subset of nodes such that
every node of the network either belongs to this subset or is
adjacent to at least one node in this set. The MDS is an important
construct if the inclusion of members into the dominating set comes
at a certain non-zero cost. For example, in the case of network
sensor placement, if placing a sensor has a non-zero cost and if
each sensor can eavesdrop on all its neighbors, then the MDS nodes
define the lowest cost placement that allows all nodes to be
monitored. Continued interest in network control, detection and
efficient spreading or curbing of network flows motivates our
interest in understanding the scaling behavior of the MDS on
stylized network models of real networks.

In particular, we focus on the properties of the MDS in scale-free
networks that are characterized by a power law degree distribution
($P(k) \sim k^{-\gamma}$). These networks constitute a class of
stylized networks which bear strong resemblance to several
real-world networks including social, infrastructural and biological
networks. Typically, values of the power law exponent $\gamma$ lie
in the range  $2 < \gamma < 3$ \cite{Broder,Jeong}, although there
are few examples of networks with $\gamma < 2$ value; for example
the co-authorship network in high-energy physics \cite{Newman}, and
some email networks \cite{Ebel}. Several algorithms have been
developed and used in previous works to generate scale-free
networks. Most of these methods are more general solutions to the
problem of constructing a network from any prescribed degree
distribution \cite{Molloy,conf}, applied to a power-law degree
distribution.

The mathematical literature focusing on bounds on dominating sets is
vast \cite{Haynes_1998}. In most prior works (with the exception of
\cite{japanese} to be discussed below), the MDS has not been studied
systematically for scale-free networks over a significantly varying
range of $\gamma$. For example, Cooper et al. \cite{Cooper} studied
the behavior of MDS size on the special class of scale-free networks
generated by preferential attachment \cite{Barabasi} (corresponding
to $\gamma=3$), and found that minimum dominating sets as well
as minimum $h$-dominating sets (where every node needs to be
dominated at least $h$ times) have sizes that are bounded above and
below by functions linear in $N$, where $N$ is the number of nodes
in the network. Other studies have focussed on MDS sizes for random
regular graphs and Erd\H{os}-R\'enyi (ER) \cite{ER} graphs. Zito
\cite{Zito} studied the size of the minimum independent dominating
set on r-regular random graphs (with $3 \leq r \leq 7$) and showed
that the size of this set (and therefore the MDS) is upper bounded
by a linear function of $N$. Recently, B\'ir\'o et al.
\cite{Biro_2012} improved the pre-factor of the $O(N)$ bound of the
size of the MDS in {\it r}-regular graphs based on a greedy
algorithm \cite{Alon_2000,Alon_1990,Arnautov_1974,Clark_1998}.
Wieland et al. \cite{Wieland} derived general bounds for {\em dense}
ER graphs (with fixed edge probability), showing that the MDS size
scales as $\log N$ (with no direct applicability to sparse graphs
with fixed average degree).

A recent study \cite{japanese} has focused on the behavior of the
MDS size on model scale-free networks with varying degree exponent,
as well as empirical networks. The authors employed the Havel-Hakimi
algorithm \cite{Havel_1955,Hakimi_1962} with random (Monte-Carlo)
edge swaps \cite{hhmc} (HHMC) for constructing synthetic networks,
and they used a binary integer programming method to obtain the MDS.
They reported that the MDS size decreases as $\gamma$ is lowered,
making heterogeneous networks very easy to control.  However, our
study of a variety of scale-free network families suggests a more
complicated picture. In particular, we find that details of the
network generation process, and the choice of the maximum degree
cutoff, bear an enormous influence on the dependence of MDS size on
$\gamma$, even when the average degree $\langle k \rangle$ is kept
fixed. The latter constraint is motivated by the need of comparing
networks (from the MDS perspective) with the same amount of
``resources", i.e., fixed average edge-cost per node, or
equivalently, fixed average degree in unweighted networks.
Naturally, for $\gamma<2$ and fixed average degree, there is
only a finite (but large) parameter range in terms of $\gamma$ and
$N$ with realizable networks.
Nevertheless, motivated by the existence of several such real-world
sparse networks \cite{Marsili_PRE2006,snap}, we also investigate
networks from various ensembles in this regime and, in particular,
how easy or hard it is to dominate them.
On the other hand, when keeping the minimum degree fixed in this
regime, the number of edges increase faster than the number of
nodes, and those networks are becoming inherently easy to dominate.
For $\gamma>2$ keeping the average or the minimum degree fixed
are equivalent constraints in the large network-size limit.

Finding the exact MDS is one of the well-known NP-hard problems of
graph theory. While it has been proven that finding a sublogarithmic
approximation to the MDS is also NP-hard, a logarithmic
approximation of the MDS \cite{raz_safra} can be found by a simple
greedy algorithm \cite{Alon_2000,Alon_1990,Arnautov_1974}. The
greedy algorithm is ideal for practical applications as it provides
a logarithmic approximation for the MDS, and runs in time linear in
the number of edges. This is certainly superior in comparison to
binary integer programming-based methods (used in \cite{japanese})
that have unknown (network structure dependent) polynomial run time,
exponential run time in the worst case, and yield no significantly
better approximation to the MDS size in finite waiting time than the
greedy algorithm, according to our experiments.

In practice, one could imagine a scenario where rather than
dominating all nodes of the network, it is sufficient to dominate
some (large) fraction of nodes. This reduces to the problem of
finding a {\it partial} minimum dominating set (pMDS) \cite{Kneis}
which is the smallest subset of nodes (and possibly a subset of the
full MDS) such that at least some given fraction of the nodes are
either in the set or adjacent to a node in this set. We investigate
the scaling of both the MDS as well as the pMDS with respect to the
network size $N$.

\section*{Results}

We start with a short description of the three network construction methods that we utilize to generate three classes of networks with the same power-law degree distribution, and a predefined average degree. Each method is a general degree sequence sampling method applied to degree sequences drawn from discrete power-law distributions with predefined parameters. The degree sequence is treated as a list of stubs (half-edges) for each node; pairs of stubs are connected to form edges.
The methods are identified by four-letter abbreviations, and are as follows:
\begin{itemize}
\item Configuration model \cite{conf,Molloy} (CONF networks), where we randomly pick any two edge stubs and connect them, until there are no more stubs to connect. This results in a multigraph; we reduce multiple links to single links, and remove self loops, to get a simple graph.
\item a Markov chain Monte Carlo method \cite{hhmc} (HHMC networks), where we first build a simple graph deterministically by the Havel-Hakimi algorithm, and then we randomize the links by swapping pairs of edges. The number of the attempted edge swaps is four times the number of edges in the network (see Supplementary Information Sections S.1 and S.4).
\item a sequential algorithm that generates samples from all possible realizations of a given degree sequence \cite{dktb} (DKTB networks, named after the authors).
      The DKTB graph-construction algorithm is based on the underlying theorems proven in Ref.~\cite{Kim_JPhysA2009}.
\end{itemize}

We use two possible subclasses of each network class, according to
the maximum degree cutoff $k_{\max}$. Either there is no explicit
cutoff, having $k_{\max}=N-1$ (where $N$ is the network size), or we
use a structural cutoff $k_{\max} = \sqrt{N}$, resulting in
uncorrelated scale-free networks
\cite{uncorrel,conf-scalefree-boguna}. When we have the $\sqrt{N}$
cutoff, we indicate it in the name of network type as cCONF, cDKTB,
or cHHMC, where \textit{c} stands for \textit{cutoff}.

As indicated in the results and in our figures, the average degree
of each individual network is kept fixed at a predetermined value
throughout all samples of each dataset. Details on controlling the
average degree are included in the Methods section, technical data
is provided in the Supplementary Information, Section S.3.

Results shown in the next subsections are generated by running a
sweep of network size $N$ and power-law exponent $\gamma$ values,
generating hundreds of network realizations, and averaging MDS size
among them for each parameter combination. The MDS of each
individual network is found by a greedy algorithm.

\subsection*{MDS Scaling With Network Size}

\begin{figure}[tbh]
\centerline{\includegraphics[width=180mm]{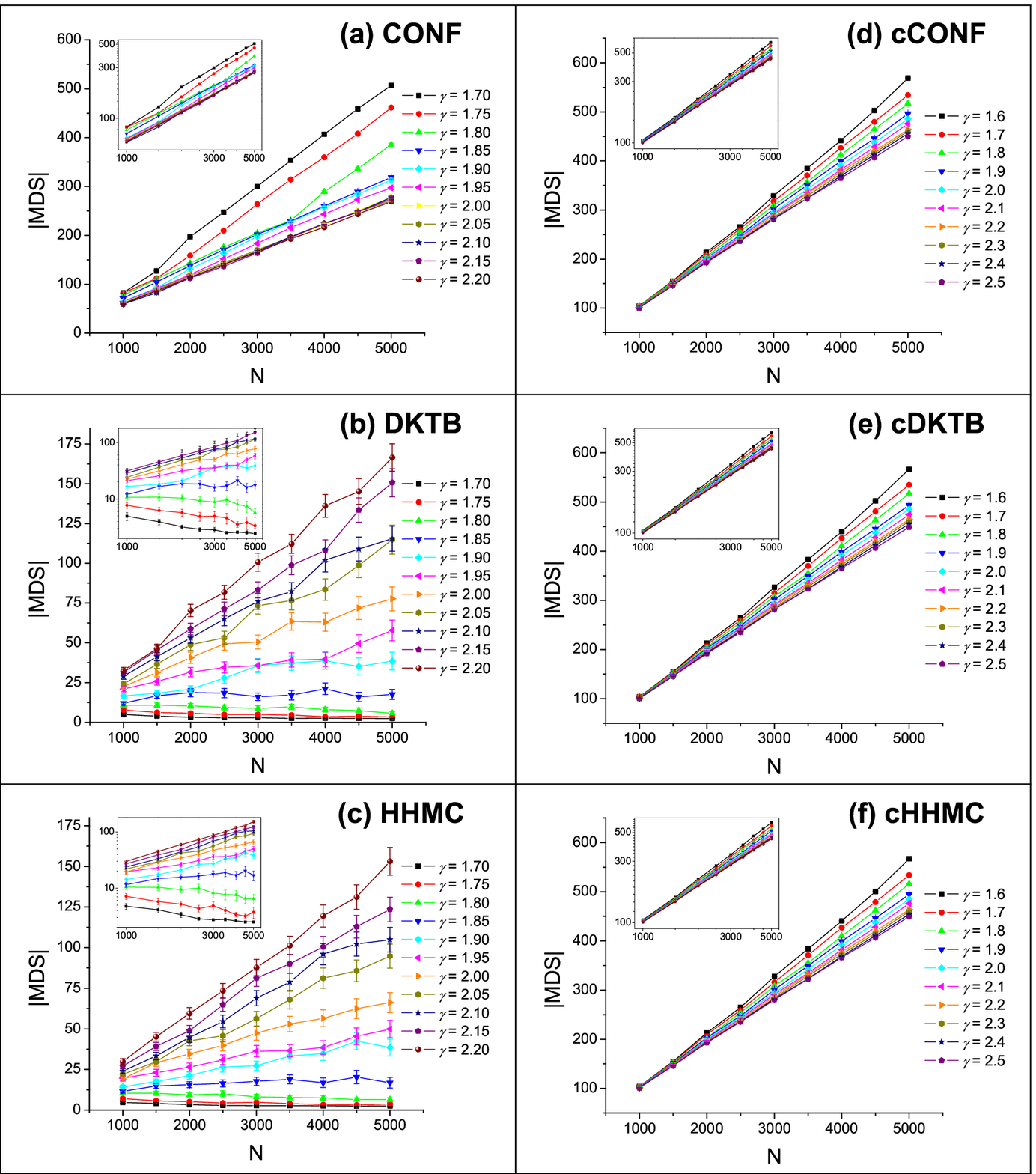}}
\caption{The size of MDS scaling with $N$, $\langle k \rangle = 14$, for all network types,
averaged over $400$ network realizations with 5 greedy searches for each at every data point.
The figure insets show the same data on log-log scales. Error bars are shown for all data points
(however, they may be very small and hidden by the larger symbols).}
\label{fig-mds-n}
\end{figure}

Figures \ref{fig-mds-n}(a)--(c) show the MDS size for networks
without any explicit upper cutoff on degree ($k_{\max}=N-1$). For CONF networks, the
MDS size scales linearly with $N$ for all $\gamma$ values. In
striking contrast, DKTB networks and HHMC networks show a marked
transition in the scaling behavior at $\gamma_c \approx 1.9$. For
$\gamma > \gamma_c$, MDS size scales linearly with $N$, whereas for
$\gamma < \gamma_c$, the MDS size appears to lose its dependence on
$N$ in the asymptotic limit. Figures \ref{fig-mds-n}(a)--(c) focus
on a subset of all considered $\gamma$ values, which range from
$\gamma = 1.6$ to $3.00$, to clearly show the scaling transition for
DKTB and HHMC networks.

Figures \ref{fig-mds-n}(d)--(f) show in contrast that with the
structural degree cutoff, ($k_{\max}=\sqrt{N}$), for all network
classes the MDS size scales linearly with $N$ irrespective of the
$\gamma$ value.

\begin{figure}[h!]
\centerline{\includegraphics[width=180mm]{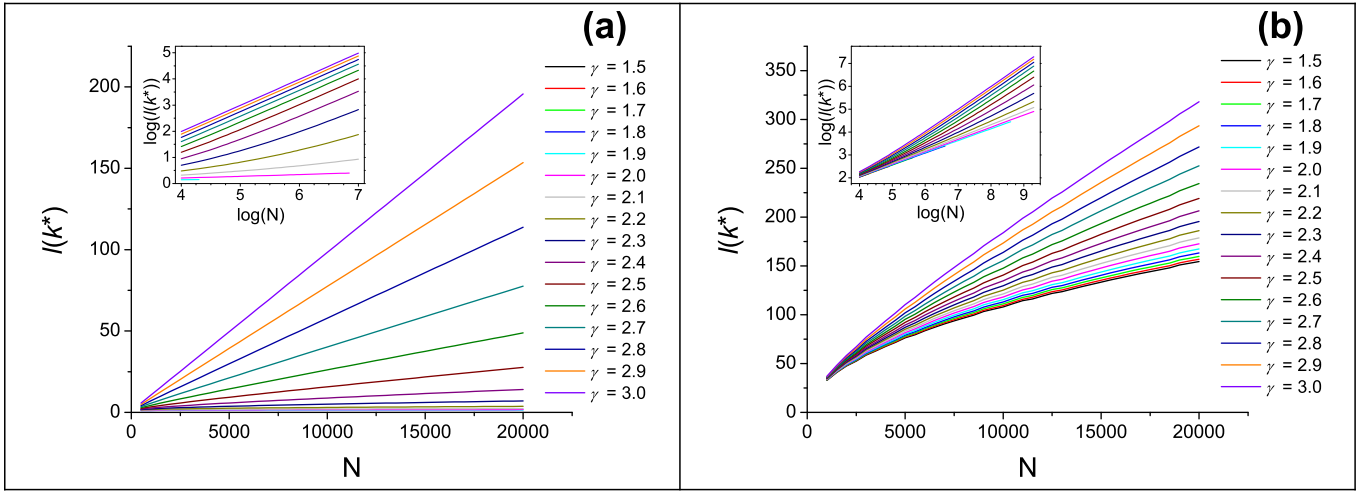}}
\caption{The scaling of the calculated lower bound of MDS size in
power-law distributions, for various power-law exponents,
$\langle k\rangle=10$. (a) $k_{\max}=N-1$, (b) $k_{\max}=\sqrt{N}$.
Figure insets show $l(k^{*})$ bounds on log-log scales. See Supplementary Information (S.7) for details.}
\label{fig-lower-bound}
\end{figure}

To better understand this scaling, we can derive a lower bound for the MDS size by considering a ``best case scenario'' for dominating the network. We use the continuous probability density function $f_{K}(k)$ equivalent to the discrete degree distribution and we define $l(k')$ as the expected number of nodes above a certain degree threshold $k'$:
\begin{equation}
l(k') = \int_{k'}^{k_{\max}} \! N f_{K}(k) \, \mathrm{d}k.
\label{eq-lower-mds}
\end{equation}
In the ``best case'', the neighborhoods of these nodes are disjoint sets (not overlapping), and therefore each node with degree $k$ dominates $k+1$ nodes (itself, and its neighbors). Then we can find the appropriate degree threshold to ensure the domination of all nodes by the following:
\begin{equation}
k^{*}:=\text{max} \left\{ k': \int_{k'}^{k_{\max}} \! N (k+1) f_{K}(k)  \, \mathrm{d}k \geq N \right\}.
\label{eq-lower-limit}
\end{equation}
Therefore, $l(k^{*})$ sets a lower bound for the size of MDS. Note,
that these formulae can be used with any degree distribution, and
$k^{*}$ can always be found numerically. Figure
\ref{fig-lower-bound} shows the $l(k^{*})$ bounds for power-law
distributions as a function of $N$ with $\langle k\rangle=10$.

There are multiple consequences of the lower bound's scaling. For
$k_{\max}=N-1$ networks, the possibility of an $O(N)$-to-$O(1)$
transition of MDS size is supported by $l(k^{*})$
(Fig.~\ref{fig-lower-bound}): it exhibits an $O(1)$ behavior for
$\gamma<2$, while it progresses to a linear scaling with
$N$ for $\gamma>2$ [Fig.~\ref{fig-lower-bound}(a)], similar to the results of DKTB and HHMC
networks. For networks with $k_{\max}=\sqrt{N}$, $l(k^{*}) \sim
\sqrt{N}$ when $\gamma<2$ and $l(k^{*}) \sim N$ when $\gamma>2$.
Note, however, that the convergence to the asymptotic behavior is
extremely slow for $2<\gamma<3$ [see insets of
Figs.~\ref{fig-lower-bound}(a) and (b)]. Thus, for the case of structural
cutoff, the lower bound indicates that the MDS size can never become
$O(1)$ and we cannot expect a sharp scaling transition. Derivation
of the asymptotic behavior of $l(k^*)$ is included in the
Supplementary Information (Section S.7).




\subsection*{Scaling of Partial Dominating Sets}

\begin{figure}[h!]
\centerline{\includegraphics[width=180mm]{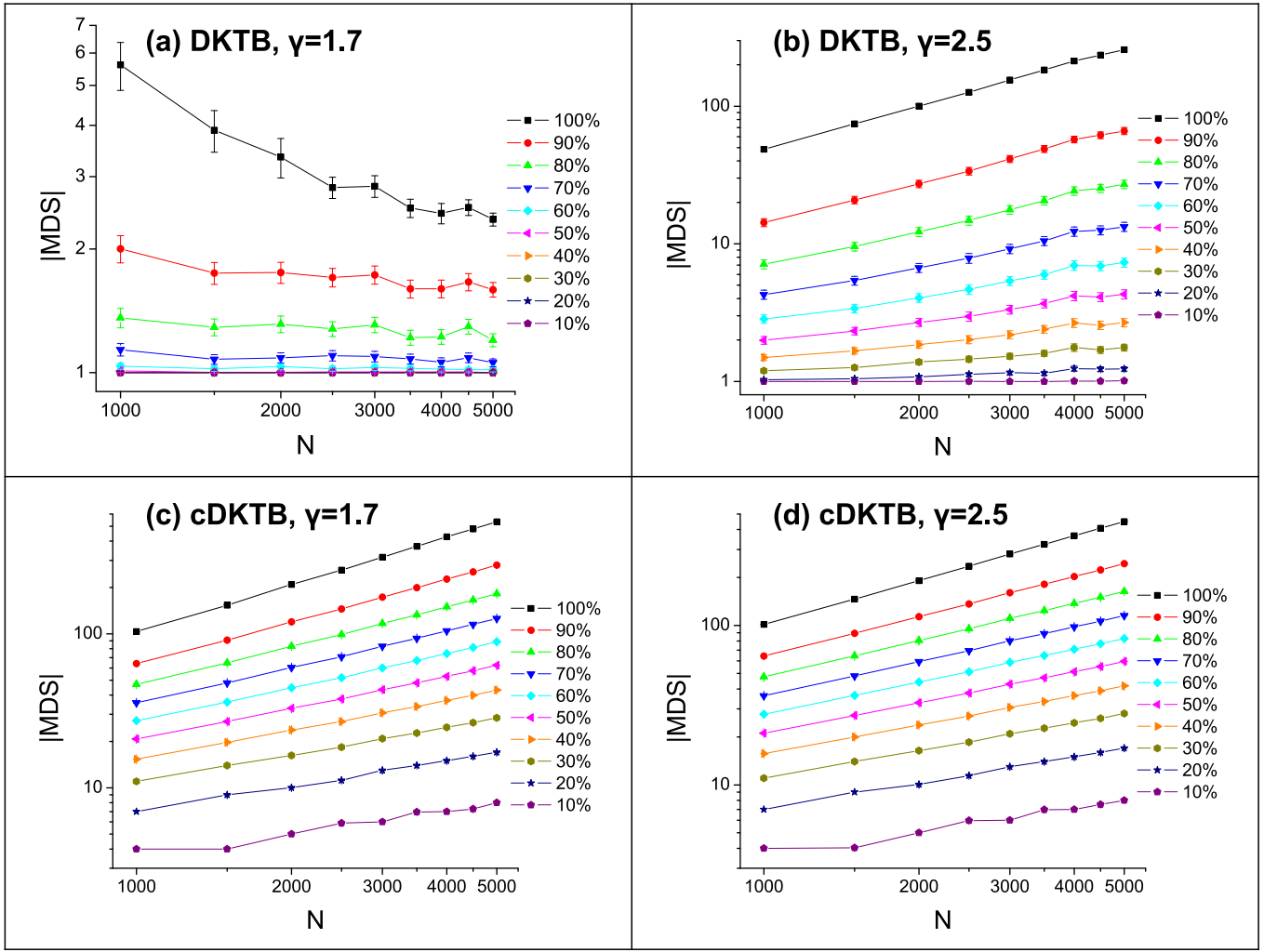}}
\caption{The size of partial MDS scaling with $N$, $\langle k \rangle = 14$,
averaged over $400$ network realizations and 5 greedy searches for each,
(a) DKTB, $\gamma=1.7$, (b) DKTB, $\gamma=2.5$, (c) cDKTB, $\gamma=1.7$, (d) cDKTB, $\gamma=2.5$.
The dominated fraction of nodes is expressed as percentage of the network size.
Error bars are shown for all data points (however, they may be very small and hidden by the larger symbols).}
\label{fig-pmds-n}
\end{figure}

Next, we study the scaling behavior of the partial MDS size with $N$
as we vary the value of the required dominated fraction $z$. In
Figure \ref{fig-pmds-n}, we present results for the DKTB class of
networks; our findings are qualitatively similar for CONF and HHMC
network classes, and networks with $k_{\max}=\sqrt{N}$. Results for
these networks are shown in Supplementary Information (Figs. S7 and
S8).

Below a certain value of $z = \widehat{k}_{\max}/N$, where
$\widehat{k}_{\max}$ is the highest realized degree in the network,
the pMDS trivially contains only the node with highest degree. Apart
from this trivial case, for $z > \widehat{k}_{\max}/N$, the size
of the pMDS exhibits the same scaling as the full MDS in the
different $\gamma$ regimes (Fig.~\ref{fig-pmds-n}). In other words,
DKTB and HHMC networks display a transition in the scaling behavior
of pMDS size from linear dependence on $N$ to virtually no
dependence on $N$ at $\gamma \approx 1.9$, while CONF networks
always show a linear scaling of pMDS size with $N$.

\begin{figure}[h!]
\centerline{\includegraphics[width=180mm]{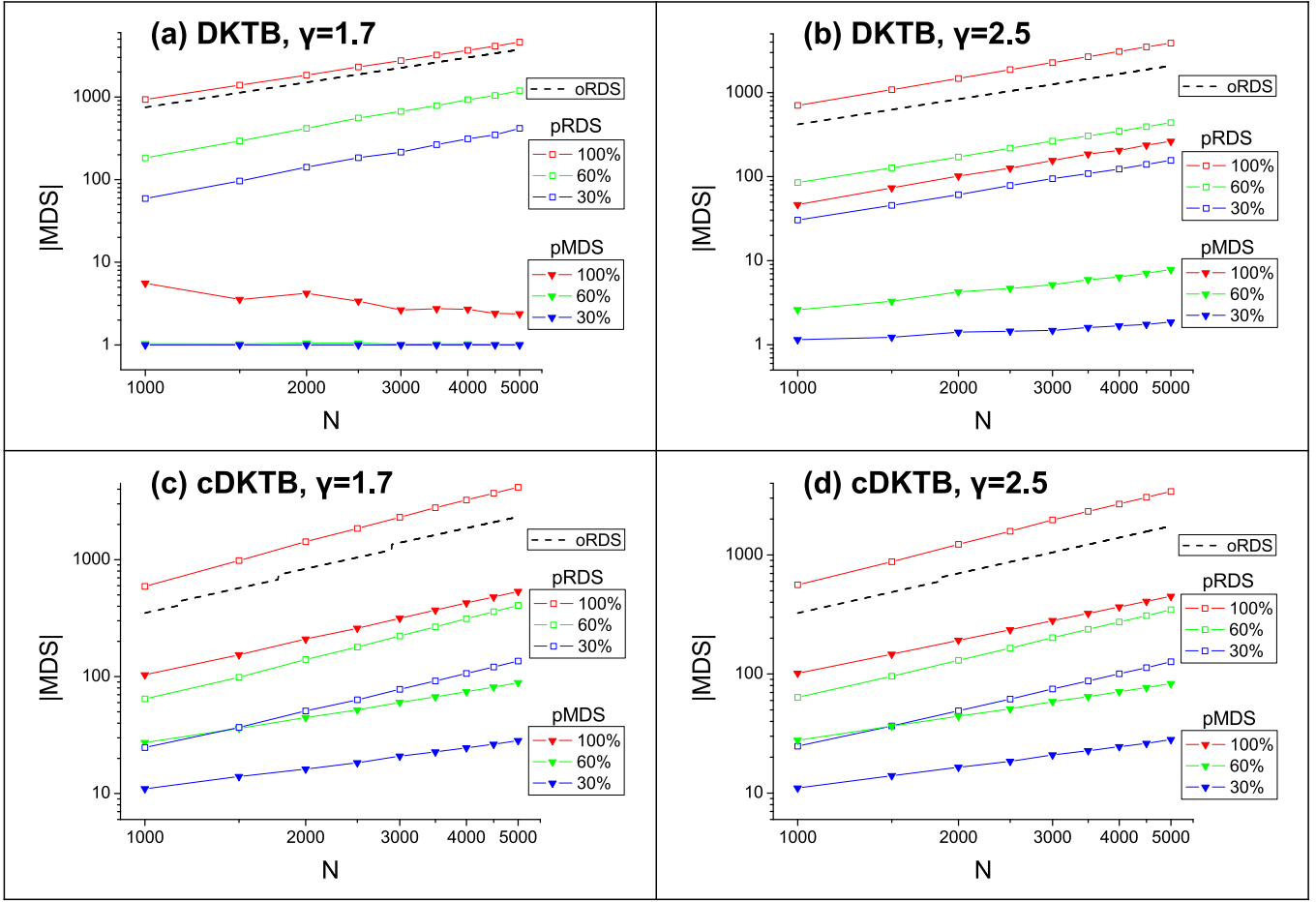}}
\caption{Comparison of partial MDS and partial RDS scaling with $N$
for DKTB (a,b) and cDKTB (c,d) networks (without and with structural cutoff, respectively); $\langle k \rangle = 14$, averaged
over $400$ network realizations and 5 greedy searches for each.
The dominated fraction of nodes is expressed as a percentage of the network size. For reference, we also show the
upper bound (dashed lines) for an optimized random dominating set
(oRDS) \cite{Haynes_1998,Alon_2000} (see text).}
\label{fig-prds-n}
\end{figure}

For a baseline-comparison to the partial MDS size obtained by greedy
algorithm, we also study the expected number of nodes needed to
dominate a given fraction of the network using random node
selection, giving a partial {\it random} dominating set (pRDS). We
run the random search five times on each realization to obtain an
expected RDS size. Figure~\ref{fig-prds-n} compares pRDS with pMDS,
showing that a simple random node selection gives approximately an
order of magnitude larger dominating set than the greedy method.
Note also, that in order to reach full domination using random node
selection, we would need to include almost all nodes of the network
in the dominating set. Further, for reference, in
Fig.~\ref{fig-prds-n} we also show the known upper bound, obtained
for {\em optimized random} selection of the dominating set (oRDS)
\cite{Alon_2000} but also applicable to the greedy algorithm
\cite{Haynes_1998,Alon_2000,Alon_1990}, for a graph with minimum
degree $k_{\min}$:
$|{\rm oRDS}|\leq N[1-k_{\min}(1+k_{\min})^{-1-1/k_{\min}}]$.
Note, that in our network construction schemes with {\em fixed}
average degree, $k_{\min}=k_{\min}(N,\gamma,\langle k\rangle)$,
hence the small jumps in the above bound when plotted as a function
of $N$ for fixed $\gamma$ and $\langle k\rangle$.


\subsection*{MDS Scaling With Power-Law Exponent}
\begin{figure}[h!]
\centerline{\includegraphics[width=180mm]{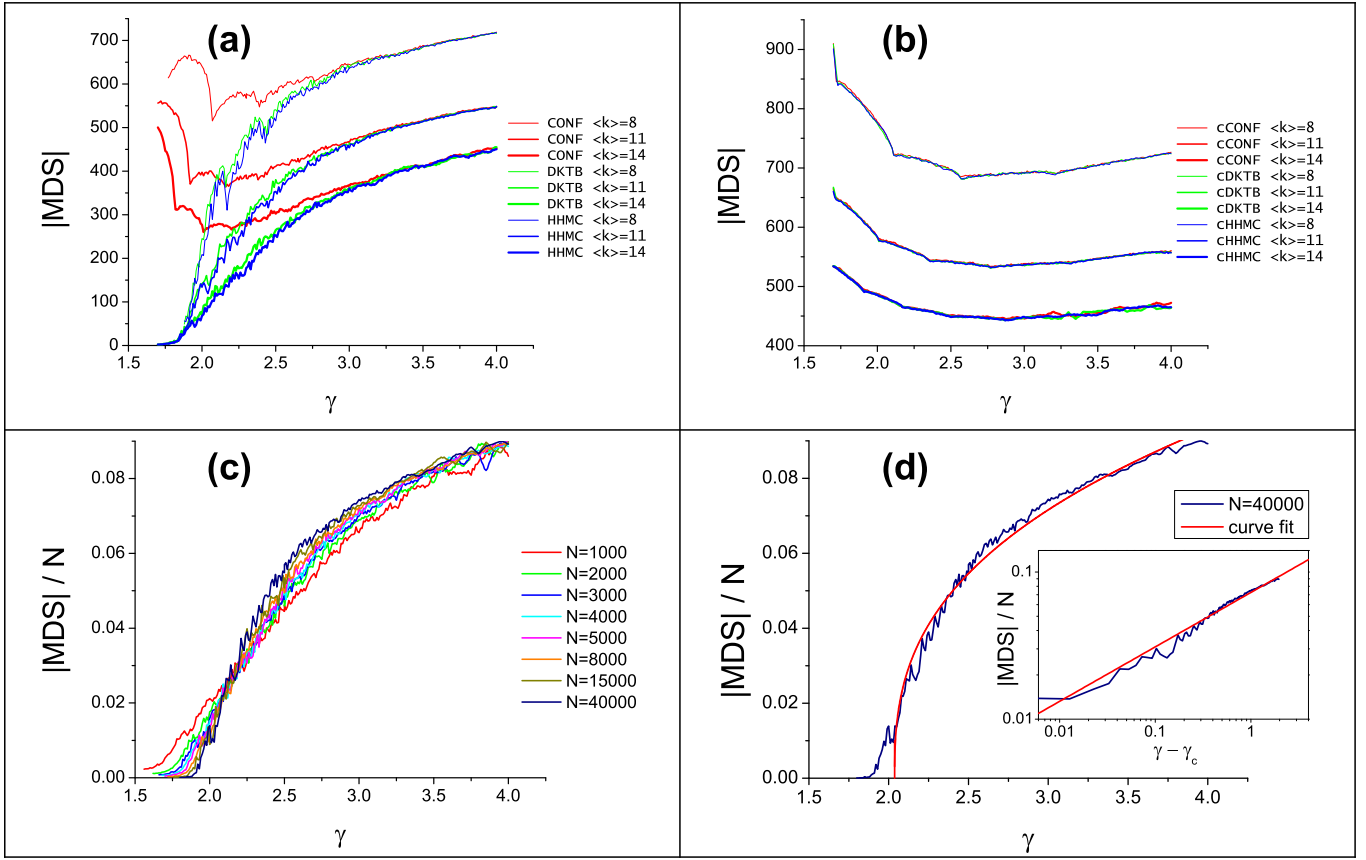}}
\caption{The size of MDS as a function of $\gamma$ for various network types and average degrees, $N=5000$,
averaged over $200$ network realizations with 5 greedy searches for each at every data point.
(a) for networks with no degree cutoff;
(b) for networks with the structural cutoff.
(c) shows the scaled MDS size vs. $\gamma$ for HHMC networks with $\langle k\rangle=14$
for various system sizes.
(d) Scaled MDS size for the largest network and the best-fit power-law (solid red curve)
in the vicinity of (and above) the transition point,
$|{\rm MDS}|/N\propto(\gamma-\gamma_c)^{\beta}$ with $\beta\approx0.37$.
Inset: same data as in (d) after shifting the horizontal axis and on log-log scales.}
\label{fig-mds-gamma}
\end{figure}

To measure the dependence of MDS size on $\gamma$, we find the MDS
for a fixed network size of $N=5000$ nodes. Results for various
$\langle k \rangle$ values are presented in Figure
\ref{fig-mds-gamma}(a) for networks with no structural cutoff, and
in Figure \ref{fig-mds-gamma}(b) for networks with a structural
cutoff.


We find a surprising trend in several cases. Perhaps, most
intriguing is the trend seen in the case of CONF networks where the
MDS size appears to have a non-monotonic dependence on $\gamma$.
Traversing increasing $\gamma$ values on a coarse scale, the MDS
size starts out large at low $\gamma$, reaches a global minimum in
the range $1.9 < \gamma < 2.3$, and then grows again as $\gamma$
increases. However, generating network samples with a finer
resolution of $\gamma$ values ($\Delta \gamma = 0.01$, reaching the
resolution of error between desired and achieved $\gamma$ values),
we also notice the existence of kinks in addition to the large scale
non-monotonicity.

By probing the dependence of MDS size on $\gamma$ for DKTB and HHMC
networks at fine resolution similar to one used for CONF, we find
only minor traces of kinks, but they are within the error margin of
MDS size. On a coarse scale, we find quantitatively similar results
for both network classes. The MDS size curve is flat at very low
values of $\gamma$, and then increases steadily beyond
$\gamma \approx 1.9$. When $\gamma > 3$, the MDS size of all three
network types converge to the same value, indicating that beyond
this point the structure of these networks are very similar
[Fig.~\ref{fig-mds-gamma}(a)].

The dependence of MDS size on $\gamma$ is strikingly different for
networks with a structural cutoff [Fig.~\ref{fig-mds-gamma}(b)]. In
this case, all three network classes show identical results for
given network parameters. For increasing $\gamma$ values the size of
MDS first decreases, then reaches its minimum at approximately $2.5
< \gamma < 3.0$, and increases again when $\gamma>3$. Since all
three classes display almost indistinguishable results, we can infer
that these networks are structurally very similar. Kinks like those
seen in CONF networks are also observed here, but with a much
smaller amplitude.

In the vicinity of (and above) the transition point, we
also found that for sufficiently large DKTB (not shown) and HHMC
[Fig.~\ref{fig-mds-gamma}(c)] networks, the scaled MDS size can be
reasonably well approximated with a power-law,
\begin{equation}
\frac{|{\rm MDS}|}{N}={\rm const.}\cdot(\gamma-\gamma_c)^{\beta} \;\,
\label{dom_tr}
\end{equation}
with $\beta\approx0.37$ [Fig.~\ref{fig-mds-gamma}(d)] (see
Discussion and Supplementary Information S.10 for further details).

\section*{Discussion}

As demonstrated clearly by the results, the specific method used for
generating a network ensemble has a profound influence on the MDS
size. This suggests that there are distinctive features in the
structures of networks generated by the different classes. From the
details of the generation methods, it might appear that DKTB and
HHMC classes produce networks that are similar in structure, and
this is certainly corroborated by our results. However, their
distinction from networks in the CONF class seems to disappear when
a structural cutoff is introduced in the degree distribution.
Although we cannot rigorously demonstrate that particular structural
features are responsible for the observed scaling behavior, we
conjecture on the origin of the distinct behaviors shown in the
Results.


It should be noted that Del Genio et al. \cite{Bassler} have shown
the non-existence of realizable graphs with a power-law degree
sequence with $0 \leq \gamma \leq 2$. However, as they point out,
their arguments refer to the situation where the prescribed degree
sequence has to be perfectly satisfied. In our methods, networks
with $1 \leq \gamma \leq 2$ are realized by removing some edge stubs
from the degree sequence that cause non-graphicality. For CONF
networks, pruning of multiple links and self loops perform this
task, while for DKTB and HHMC networks, a Havel-Hakimi-based
graphicality correction algorithm is applied (see Methods and
Supplementary Information Section S.1 for details). It is notable,
that we can choose appropriate parameters such that even after
pruning, or graphicality-preserving stub removal, we have a network
whose degree distribution (in particular, its tail)
approximates the desired power-law. (For the degree distribution of
the networks obtained this way, see Supplementary Information S.9.)
As a result of these procedures, our networks in this range of
$\gamma$ are not exact realizations of perfect power-law degree
sequences, and therefore do not contradict the fundamental results
of Ref.~\cite{Bassler}.

When the structural cutoff is not imposed on the degree sequence,
the non-graphicality below $\gamma=2$ plays an important role in the
scaling transition of the MDS size with $N$. When $\gamma<2$, there
are too many edge stubs in the prescribed degree sequence, and some
of them have to be removed to resolve non-graphicality. Different
network construction methods solve this problem differently. With
respect to MDS scaling behavior, the key difference is in the
treatment of the highest degrees. In case of CONF, the formation of
multiple links is allowed during the stub connection process, and
the duplicate links are pruned later. Since the realized multiple
links predominantly connect stubs belonging to high-degree nodes
with each other \cite{conf-scalefree-boguna,wconf-serrano}, the
large degrees of the hubs are effectively wasted in connections that
do not improve their potential to dominate. Furthermore, as a
consequence, the interconnections of low degree nodes become more
dominant, forming a relatively sparse web, which necessitates the
inclusion of many nodes in the dominating set, preventing it to
become $O(1)$. However, in case of the Havel-Hakimi-based
graphicality correction (used in HHMC and DKTB methods) the stubs of
the highest degree nodes are connected first, ensuring that these
nodes are present in the network as hubs. The MDS scaling transition
can therefore be explained by the scaling of the largest realized
degree (also known as the natural cutoff of the degree sequence),
$\widehat{k}_{\max} \sim N^{\frac{1}{\gamma-1}}$
\cite{conf-scalefree-boguna,Dorogovtsev_2003}. When $\gamma < 2$,
the MDS size becomes $O(1)$ because the largest degree, and
potentially the second and third largest degrees become $O(N)$, and
the network is dominated by these nodes. In essence, we find that
the domination transition is directly related to the underlying
graphicality transition \cite{Bassler}: the same underlying
structural properties which are responsible for the graphicality
transition \cite{Bassler} in the infinite network-size limit allow
for the $O(N)$-to-$O(1)$ domination transition for large but finite
DKTB and HHMC networks. In other words, those finite DKTB and HHMC
network realizations which happen to exist for $\gamma<2$ can be
dominated by an $O(1)$ MDS. The sharp emergence of the
$O(N)$ minimum dominating set (and the existence of the domination
transition) is also supported by the power-law behavior of the
scaled MDS size just above the transition point
[Figs.~\ref{fig-mds-gamma}(d)].

The small difference between our numerically observed value of the
domination transition at around $\gamma\approx1.9$ and the
$\gamma=2$ value of the graphicality transition \cite{Bassler} might
lie in finite-size effects and in the $\log(N)$ accuracy of the
greedy algorithm with respect to the size of the true MDS
[Figs.~\ref{fig-mds-gamma}(c) and (d)].

\begin{figure}[h!]
\centerline{\includegraphics[width=180mm]{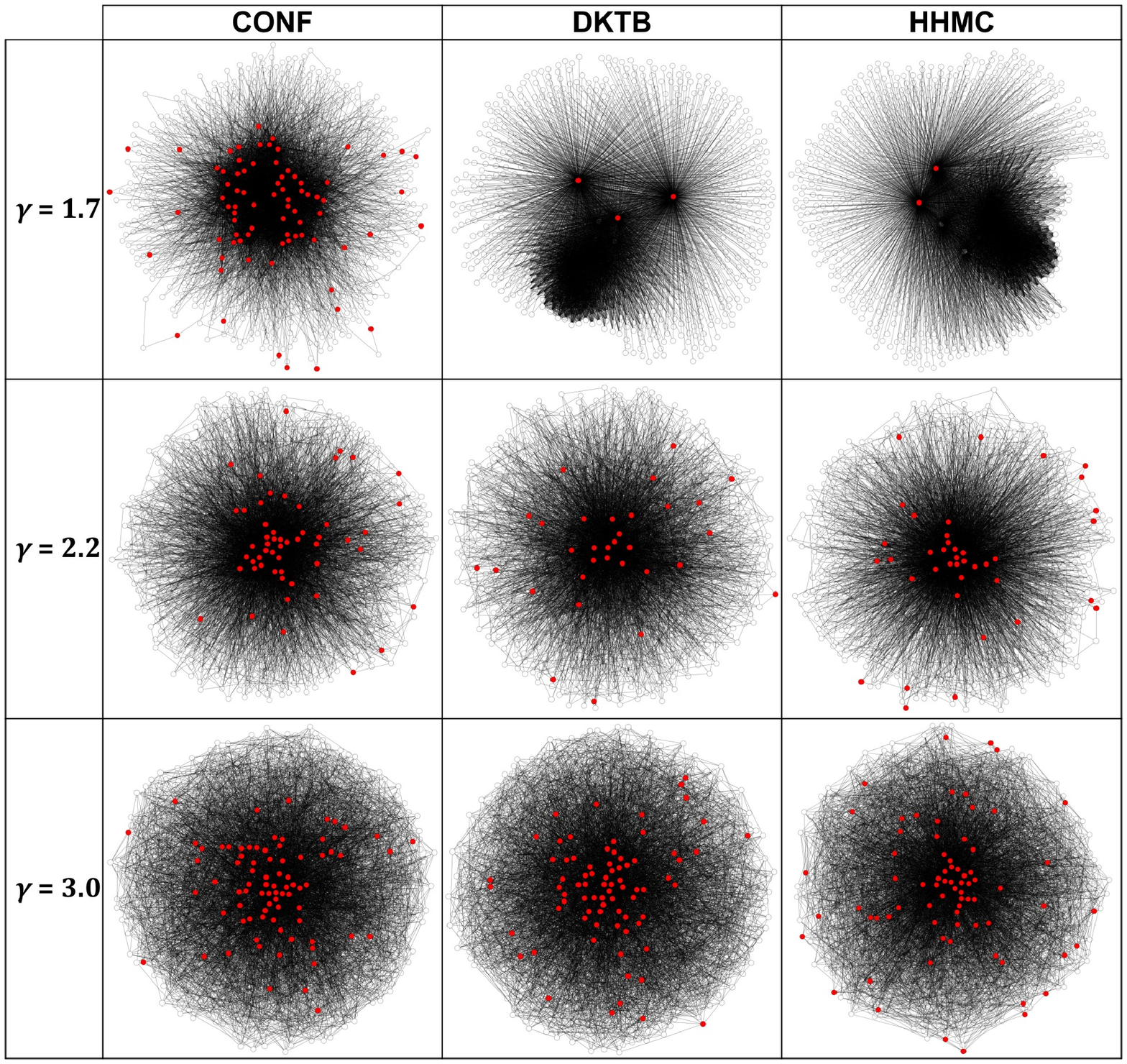}}
\caption{Visualization of typical scale-free networks of each type with $k_{\max}=N-1$ at three different power-law exponent values, embedded by the SFDP layout engine of Graphviz visualization software \cite{graphviz}. In all networks, $N=1000$ and $\langle k \rangle=14$; the colored nodes belong to the MDS.}
\label{fig-vis}
\end{figure}

\begin{figure}[t!]
\centerline{\includegraphics[width=180mm]{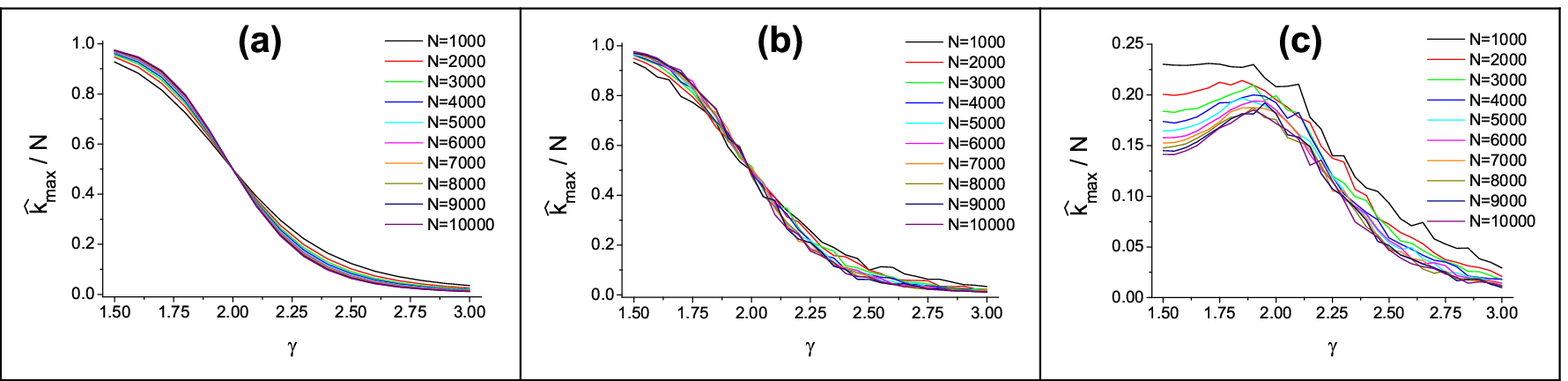}}
\caption{Scaling of maximum realized degree $\widehat{k}_{\max}$ with power-law exponent $\gamma$, for various network sizes. (a) theoretical expected value from power-law distribution, (b) degree sequence with graphicality correction (HHMC and DKTB networks), (c) degree sequence after pruning multiple links (CONF networks).}
\label{fig-kmax-gamma}
\end{figure}

The different treatment of largest degrees in different network
classes can be illustrated by plotting $\widehat{k}_{\max}$ against
$\gamma$, see Figure \ref{fig-kmax-gamma}. Note, that for the
theoretical value we need to derive and evaluate the exact formula
from the degree distribution, see Supplementary Information Section S.6.
Further, the markedly different structure of CONF networks compared
to HHMC and DKTB networks in the absence of a structural cutoff also
shows up in the network visualizations in Fig.~\ref{fig-vis}.

\begin{figure}[h!]
\centerline{\includegraphics[width=180mm]{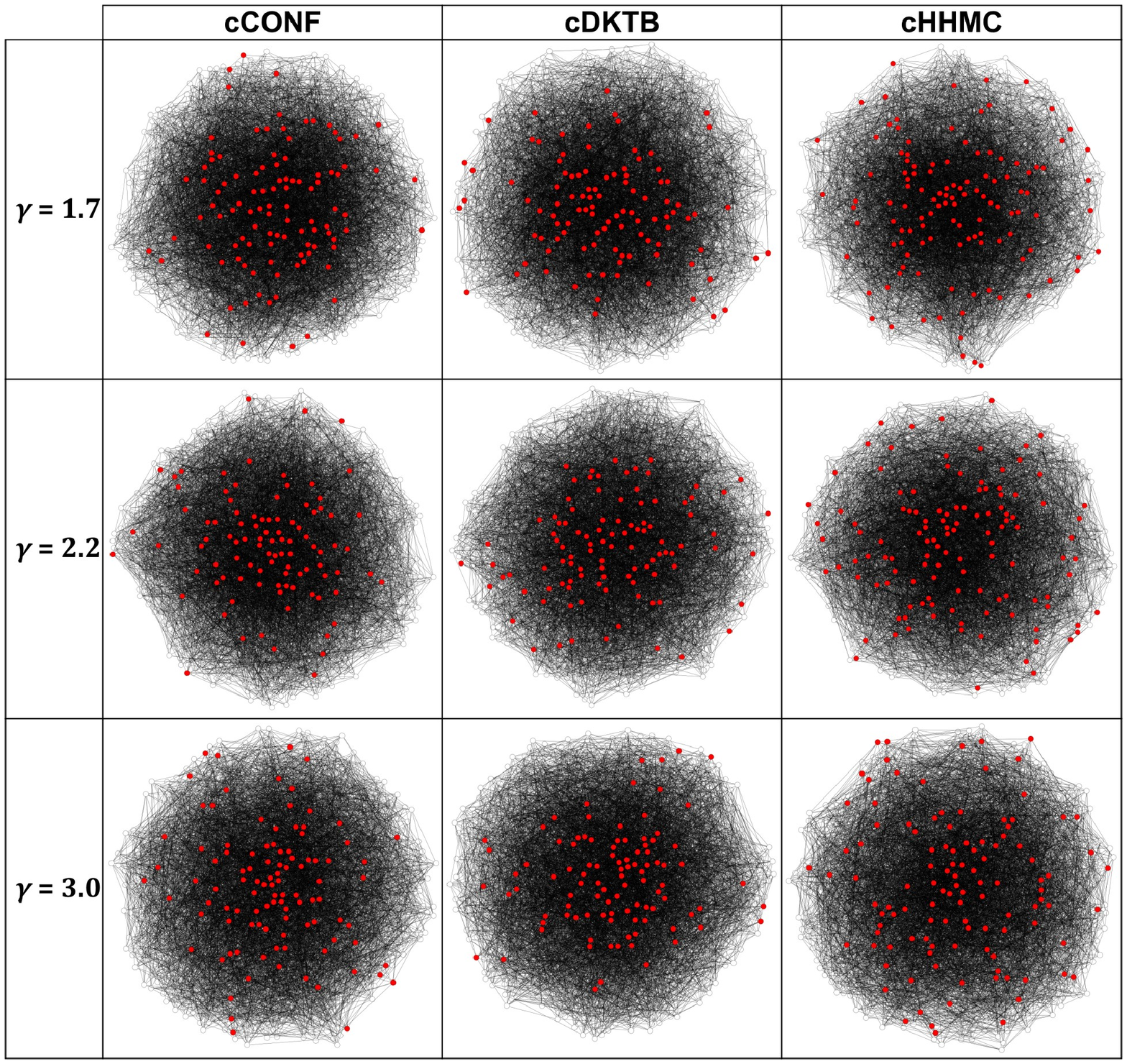}}
\caption{Visualization of typical scale-free networks of each type with $k_{\max}=\sqrt{N}$ at three different power-law exponent values, embedded by the SFDP layout engine of Graphviz visualization software \cite{graphviz}. In all networks, $N=1000$ and $\langle k \rangle=14$; the colored nodes belong to the MDS.}
\label{fig-vis2}
\end{figure}

In contrast, with the structural cutoff, networks generated using
the three different methods appear to share similar structure as can
be seen in Fig.~\ref{fig-vis2}, from the similar scaling of MDS size
with $N$, and the dependence of MDS size on $\gamma$. The
restrictive $k_{\max}=\sqrt{N}$ cutoff precludes the scaling of MDS
size from becoming $O(1)$, as shown by the $l(k^{*})$ lower bound in
Fig.~\ref{fig-lower-bound}(b).



The non-monotonic behavior of the size of the MDS with $\gamma$
(with the exception of the DKTB and HHMC constructions with no
maximum degree cutoff) are in part the consequence of the stringent
constraint of resources for domination (fixed average degree, i.e.,
fixed number of edges for fixed $N$): for small and decreasing
values of $\gamma$, while maintaining a fixed average degree for a
given network size $N$, the minimum degree decreases, and there are
$O(N)$ number of such nodes. However, in the DKTB and HHMC networks
the largest hub can have $O(N)$ links, and it has the potential
alone to connect to (and dominate) the nodes with the lowest degree,
hence the monotonic behavior with $\gamma$ (and the transition to
O(1) domination) for these networks [Fig.~\ref{fig-mds-gamma}(a)].

Kinks seen in the curves of MDS size when plotted against $\gamma$
are the result of controlling the average degree with very high
precision. Smooth change of the control parameters introduces
gradual changes in the network structure, however, the average
degree does not change smoothly (although it is monotonic; see
Supplementary Information Section S.3). Conversely, when we probe a range of
$\gamma$ values, we need a smooth control over the average degree,
requiring non-smooth changes in control parameters and hence in the
network structure. Therefore, we can expect that any
structure-dependent quantity, like the MDS size, will show kinks
with respect to $\gamma$.

The results reported by Nacher et al. \cite{japanese} suggest that
for a given $\langle k \rangle$, decreasing $\gamma$ results in a
monotonic lowering of the MDS size. However, they only studied the
HHMC method of network generation with a variable cutoff. By
introducing a well-defined structural cutoff, and in addition
studying two other classes of networks, we show that precise details
of the network construction have a strong impact on the trend in MDS
size as $\gamma$ is varied.


In summary, we have shown through extensive numerical experiments,
that the size of the minimum dominating set approximated by a greedy
algorithm undergoes a transition in its scaling with respect to $N$
only for particular methods of network construction in the absence
of a structural cutoff. For the configuration model construction, or
the other construction methods with a structural cutoff, no such
transition is observed. However, intriguingly, in the presence of a
structural cutoff, the MDS size increases as $\gamma$ is lowered
below $2$. Thus our results demonstrate that it is not sufficient to
have a scale-free network with $\gamma < 2$ to have an easily
dominated (easily controllable) network; intricate details in the
wiring of the network must also be taken into consideration.


%
%


\section*{Methods}

As the first step of network construction, we generate a discrete
power-law degree distribution, and we also calculate the cumulative
distribution function (CDF) associated with the desired discrete
power-law degree distribution. Then, using inverse sampling of the
CDF, we generate the degree sequence of the network. We find that
using a discrete distribution results in much better accuracy of the
desired power-law exponent than sampling degrees from a continuous
distribution with rounding.

Once the degree sequence is generated, each node is assumed to have
as many edge stubs as its degree. The process of network generation
connects pairs of edge stubs to form edges, the three network
construction methods (CONF, DKTB, HHMC) carry out this task
differently, according to their specific algorithms. In CONF and
DKTB, the random selection of edge stubs gives a random realization
of the degree sequence. In contrast, the HHMC method connects edge
stubs deterministically (using the Havel-Hakimi algorithm), and a
random realization is obtained by a Markov chain of swapping edge
pairs. The mixing time of this Markov chain is determined
empirically, see Supplementary Information (Section S.4).

HHMC and DKTB methods are ``exact methods'' since they do not alter
the given degree sequence while constructing the network. Therefore,
we must supply them with {\it graphical} degree sequences, i.e.
sequences for which it is assured that a simple graph realizing that
sequence exists. To ensure graphicality, we devised a
\textit{graphicality correction} method (See Supplementary
Information, Alg. 1 in S.1), based on the Havel-Hakimi algorithm. The goal
of the original algorithm is to test the graphicality of a degree
sequence. It reports failure when some stubs of a node cannot be
connected to other nodes. Instead of reporting failure, we simply
remove these stubs from the degree sequence, making the remaining
sequence graphical. This correction precedes the network
construction step for HHMC and DKTB networks, but it is not needed
for CONF networks, because every degree sequence is graphical for a
multigraph. We effectively remove any non-graphicality of the degree
sequences when we remove multiple links and self loops to create a
simple graph.

The final step of network generation is to ensure that we have a
single connected component. Further details are provided in
Supplementary Information Section S.1, including a flow diagram that illustrates
all steps of the network generation procedure.

We control the average degree of the networks by setting an
appropriate lower degree cutoff $k_{\min}$. In order to have a very
fine control over the average degree, we also remove a given
fraction $f$ of the lowest degrees from the degree distribution, $f
\in \left[0,1\right)$. To calibrate which $k_{\min}$ and $f$ values
result in which average degrees, we constructed a high-accuracy
lookup table by generating network samples for all possible
parameter combinations of $k_{\min}$, $f$, $N$ and $\gamma$, for all
network classes. By numerically inverting this table, we can find
the needed parameters to achieve any desired average degree for any
given network class. Details on this construction, the corresponding
lookup table and achieved accuracy are included in the Supplementary
Information (Section S.3).

Note, that the smallest reachable $\gamma$ is defined by $\langle k
\rangle$. The average degree of a network increases rapidly as
$\gamma$ decreases for given $k_{\min}$ (see Figure S2 in the
Supplementary Information), and for a given $\gamma$ the lowest
$\langle k \rangle$ is obtained when $k_{\min} = 1$. Consequently,
for a fixed $\langle k \rangle$, the lowest possible $\gamma$ is the
value at which the $\langle k \rangle$ vs. $\gamma$ curve for
$k_{\min}=1$ attains the desired $\langle k \rangle$. This lower
limit on $\gamma$ for each $\langle k \rangle$ can be determined
from the lookup tables. We study the MDS scaling behavior in $8 \leq
\langle k \rangle \leq 16$ range, because it allows for a wide range
of $\gamma$ values.

Since finding the MDS is NP-hard, we approximate the exact solution
by using a sequential greedy algorithm. Starting with an empty set
$\mathcal{D}$, at each step the algorithm adds that node to
$\mathcal{D}$ which yields the largest increase in the number of
dominated nodes in the network. When there are multiple candidate
nodes yielding the maximal increase in domination, the algorithm
chooses one randomly (uniformly among candidates). These steps
continue until all nodes are dominated and then the algorithm
terminates with $\mathcal{D}$ storing the approximated MDS. The
greedy algorithm yields a $1+\log N$ \cite{raz_safra} approximation
to the size of MDS, and has a time complexity of $O(E)$. See the
Supplementary Information for implementation details.

We found that the MDS sizes given by the greedy algorithm for any
single network follow approximately a Gaussian distribution, with at
least an order of magnitude smaller standard deviation than the
standard deviation of average MDS size among multiple network
samples. (See distributions and histograms obtained by
the greedy search in the Supplementary Information, section S.8.) Therefore,
for any given network, we find it sufficient to run the greedy
algorithm five times, to obtain a reliable estimate of the average
MDS size. Similarly obtained results for all network realizations
for a given combination of parameter values, and a given network
class are averaged to obtain an estimate of the mean greedily
approximated MDS size.

In order to find the pMDS for a given dominated fraction $z$, we use
the same greedy approach as for the full MDS, except that we
terminate the algorithm when the desired fraction of nodes are
dominated.

\section*{Acknowledgments}
Helpful comments on this work by K.E. Bassler, \'E. Czabarka, L. Sz\'ekely, and Z.
Toroczkai, are gratefully acknowledged.
This work was supported in part by
grant No. FA9550-12-1-0405 from the U.S. Air Force Office of Scientific Research (AFOSR)
and
the Defense Advanced Research Projects Agency (DARPA),
by the Defense Threat Reduction Agency (DTRA) Award No. HDTRA1-09-1-0049,
by the Army Research Laboratory under Cooperative Agreement Number W911NF-09-2-0053,
by the Army Research Office grant W911NF-12-1-0546, and
by the Office of Naval Research Grant No. N00014-09-1-0607.
The views and conclusions contained in this document are those of
the authors and should not be interpreted as representing the
official policies either expressed or implied of the Army Research
Laboratory or the U.S. Government.

\section*{Author Contributions}
F.M., S.S., B.K.S. and G.K. designed the research;
F.M. implemented and performed numerical experiments and simulations;
F.M., S.S., B.K.S. and G.K. analyzed data and discussed results;
F.M., S.S., B.K.S. and G.K. wrote and reviewed the manuscript.

\section*{Additional Information}
Competing financial interests: The authors declare no competing financial interests.



\vspace*{-4cm}


\begin{thebibliography}{100}





\bibitem{Eubank}
Eubank, S., Anil Kumar, V. S.,  Marathe, M. V., Srinivasan, A., Wang N.
Structural and algorithmic aspects of massive social networks.
In SODA '04 Proc. of the fifteenth annual ACM-SIAM symposium on Discrete algorithms, 718--727 (2004).

\bibitem{Cowan}
Cowan,  N. J., Chastain,  E. J., Vilhena, D. A., Freudenberg, J. S., Bergstrom, C. T.
Nodal Dynamics, Not Degree Distributions, Determine the Structural Controllability of Complex Networks.
PLoS ONE 7(6): e38398 (2012).

\bibitem{Kelleher_1988}
Kelleher, L., Cozzens, M.
Dominating Sets in Social Network Graphs.
Math. Soc. Sciences 16, 267--279 (1988).

\bibitem{Wang}
Wang, F. {\it et al.}
On positive influence dominating sets in social networks.
Theo. Comp. Sci. 412, 265--269 (2011).

\bibitem{Broder}
Broder, A. {\it et al.}
Graph structure in the web.
Computer Networks: The International Journal of Computer and Telecommunications Networking 33, 309--320 (2000).

\bibitem{Jeong}
Jeong, H., Mason, S., Barabasi, A.-L., Oltvai, Z. N.
Lethality and centrality in protein networks.
Nature 411, 41--42 (2001).

\bibitem{Newman}
Newman, M. E. J.
Scientific collaboration networks: I. Network construction and fundamental results.
Phys. Rev. E 64, 016131 (2001).

\bibitem{Ebel}
Ebel, H., Mielsch, S., Bornholdt, S.
Scale-free topology of e-mail networks.
Phys. Rev. E 66, 035103(R) (2002).

\bibitem{Molloy}
Molloy, M., Reed, B.
A Critical Point for Random Graphs with a Given Degree Sequence.
Random Structures and Algorithms 6, 161--180 (1995).

\bibitem{conf}
Britton, T., Deijfen, M., Martin-L\"{o}f, A.
Generating simple random graphs with prescribed degree distribution.
J. Stat. Phys. 124, 1377--1397 (2005).

\bibitem{Haynes_1998}
Haynes, T.W., Hedetniemi, S.T.,  Slater, P.J.
Fundamentals of Domination in Graphs (Marcel Dekker, New York, 1998).

\bibitem{japanese}
Nacher, J. C., Akutsu, T.
Dominating scale-free networks with variable scaling exponent: heterogeneous networks are not difficult to control.
New J. Phys. 14, 073005 (2012).

\bibitem{Cooper}
Cooper, C., Klasing, R., Zito, M.
Lower bounds and algorithms for dominating sets in web graphs.
Internet Mathematics 2, 275--300 (2005).

\bibitem{Barabasi}
Barab\'asi, A.-L., Albert R.
Emergence of scaling in random networks.
Science 286, 509--512 (1999).

\bibitem{ER}
Erd\H{o}s, P., R\'enyi A.
On the evolution of random graphs.
Publ. Math. Inst. Hung. Acad. Sci. {\bf 5}, 17--61 (1960).

\bibitem{Zito}
Zito, M.
Greedy Algorithms for Minimisation Problems in Random Regular Graphs.
ESA '01 Proceedings of the 9th Annual European Symposium on Algorithms, 525--536 (2001).

\bibitem{Biro_2012}
B\'ir\'o, Cs., Czabarka, \'E., Dankelmann, P., Sz\'ekely, L.
Bulletin of the I. C. A., 64, 73--82, (2012).

\bibitem{Alon_2000}
Alon, N., Spencer, J.H.
The Probabilsitic Method.
2nd ed. (Willey, New York, 2000).

\bibitem{Alon_1990}
Alon, N.
Transversal numbers of uniform hypergraphs.
Graphs Combin., 6, 1--4 (1990).

\bibitem{Arnautov_1974}
Arnautov, V.I.
Estimation of the exterior stability number of a graph by means of the minimal degrees of the vertices.
Prikl. Mat. i Programmirovanie 11, 3--8, 126, 1974 (in Russian).

\bibitem{Clark_1998}
Clark, W.E., Shekhtman, B., Suen, S., Fisher., D.C.
Upper bounds for the domination number of a graph.
Congr. Numer., 132, 99--123 (1998).

\bibitem{Wieland}
Wieland B., Godbole A. P.
On the Domination Number of a Random Graph.
The Electronic Journal of Combinatorics 8, R37 (2001).


\bibitem{Havel_1955}
Havel, V.
A remark on the existence of finite graphs.
Casopis Pest. Mat. 80, 477--480 (1955).

\bibitem{Hakimi_1962}
Hakimi, S.L.
On the realizability of integers as the degrees of the vertices of a linear graph - I.
J. SIAM Appl. Math. 10, 496--506 (1962).

\bibitem{hhmc}
Viger, F., Latapy, M.
Efficient and simple generation of random simple connected graphs with prescribed degree sequence.
In: The 11th Intl. Comp. and Combin. Conf., 440--449 (2005).


\bibitem{Marsili_PRE2006}
Seyed-allaei, H., Bianconi, G. and Marsili, M.
Scale-free networks with an exponent less than two.
Phys. Rev. E 73, 046113 (2006).

\bibitem{snap}
Stanford Network Analysis Project (SNAP),
http://snap.stanford.edu/data (Accessed April 2, 2013).
%

\bibitem{raz_safra}
Raz, R., Safra, S.
A Sub-Constant Error-Probability Low-Degree Test, and a Sub-Constant Error-Probability PCP Characterization of NP.
In Proc. of the 29th annual ACM Symposium on Theory of Computing, 475--484 (1997).

\bibitem{Kneis}
Kneis, J., Molle, D., Rossmanith, P.
Partial vs. Complete Domination: t-Dominating Set.
In SOFSEM 2007, Theory and Practice of Computer Science, Lecture Notes in Computer Science, 4632, 367--376 (2007).

\bibitem{dktb}
Del Genio, C. I., Kim, H., Toroczkai, Z., Bassler, K. E.
Efficient and Exact Sampling of Simple Graphs with Given Arbitrary Degree Sequence.
PLoS ONE 5(4): e10012 (2010).

\bibitem{Kim_JPhysA2009}
Kim, H., Toroczkai, Z., Erd\H{o}s, P.L., Mikl\'os, I., and Sz\'ekely, L.A.
Degree-based graph construction. J. Phys. A: Math. Theor. 42, 392001 (2009).

\bibitem{uncorrel}
Catanzaro, M., Boguna, M., Pastor-Satorras, R.
Generation of uncorrelated random scale-free networks.
Physical Review E 71, 027103 (2005).

\bibitem{conf-scalefree-boguna}
Bogu\~{n}\'{a}, M., Pastor-Satorras, R., Vespignani, A.
Cut-offs and finite size effects in scale-free networks.
Eur. Phys B 38, 205--509 (2004).

\bibitem{Bassler}
Del Genio, C. I., Gross, T., Bassler, K. E.
All Scale-Free Networks Are Sparse.
Phys. Rev. Lett. 107, 178701 (2011).

\bibitem{wconf-serrano}
Serrano, \'{A}. M., Bogu\~{n}\'{a}, M.
Weighted Configuration Model.
AIP Conf. Proc. 776, 101--107 (2005).

\bibitem{Dorogovtsev_2003}
Dorogovtsev, S.N., Mendes, J.F.F.
Evolution of networks. Adv. Phys. 51, 1079--1187 (2002).

\bibitem{graphviz}
Graphviz --- Graph Visualization Software, www.graphviz.org (Accessed April 2, 2013).


\end{thebibliography}
\end{document}